\documentclass[12pt,english,floatfix,superscriptaddress,aps,preprint,showpacs]{revtex4}
\usepackage{amsmath}
\usepackage{amssymb}
\usepackage{amsbsy}
\usepackage{amsfonts}
\usepackage{amsopn}
\usepackage{amstext}
\usepackage{graphicx}
\usepackage{esint}
\usepackage{amssymb}
\usepackage{amsfonts}
\usepackage{amsmath}
\usepackage{graphicx}
\usepackage[english]{babel}
\usepackage{color}
\usepackage[dvips]{epsfig}
\usepackage[dvips]{graphicx}
\usepackage{float}
\usepackage{units}
\usepackage{textcomp}
\usepackage{babel}

\begin{document}

\title{Matter-gravity scattering in the presence of spontaneous Lorentz
violation}
\author{R. V. Maluf}
\email{r.v.maluf@fisica.ufc.br}
\affiliation{Universidade Federal do Cear\'a (UFC), Departamento de F\'isica, Campus do
Pici, Fortaleza - CE, C.P. 6030, 60455-760 - Brazil}
\author{Victor Santos}
\email{victor_santos@fisica.ufc.br}
\affiliation{Universidade Federal do Cear\'a (UFC), Departamento de F\'isica, Campus do
Pici, Fortaleza - CE, C.P. 6030, 60455-760 - Brazil}
\author{W. T. Cruz}
\email{wilamicruz@gmail.com}
\affiliation{Instituto Federal de Educa\c{c}\~ao, Ci\^encia e Tecnologia do Cear\'a (IFCE),
Campus Juazeiro do Norte, 63040-000 Juazeiro do Norte - CE - Brazil }
\author{C. A. S. Almeida}
\email{carlos@fisica.ufc.br}
\affiliation{Universidade Federal do Cear\'a (UFC), Departamento de F\'isica, Campus do
Pici, Fortaleza - CE, C.P. 6030, 60455-760 - Brazil}
\date{\today}

\begin{abstract}
Considering quantum gravity within the framework of effective field theory,
we investigated the consequences of spontaneous Lorentz violation for the
gravitational potential. In particular, we focus our attention on the
bumblebee models, in which the graviton couples to a vector $B_{\mu}$ that
assumes a nonzero vacuum expectation value. The leading order corrections
for the nonrelativistic potential are obtained from calculation of the
scattering matrix of two scalar particles interacting gravitationally. These
corrections imply anisotropic properties associated with the bumblebee background and also add a Darwin-like term for Newton\textquoteright{s} potential.
\end{abstract}


\maketitle

\section{Introduction}

A longstanding problem in theoretical physics is the conciliation between
the Standard Model (SM) describing the behavior of elementary particles and
General Relativity (GR), which accounts the large scale physics dominated by
gravity. With such a conciliation, both theories, which are
extremely well tested, should appear as low-energy descriptions of
a single and fundamental (and yet unknown) theory of quantum
gravity. This framework opens the possibility for the discovery of new
phenomena, not described by any of these effective theories. Unfortunately, since quantum gravity effects are relevant at energy scales of the order of the Planck mass $m_{\mbox{P}}\sim 1.22\times 10^{19}\mbox{\ GeV}$, no experimental evidence for the signature of
a more fundamental physics has been obtained up to now.

Despite the fact that Planck scale dynamics remains impossible to access experimentally, a great deal of work has been performed by exploring the point of view that quantum gravity phenomena can be observed by amplification of its effects at attainable energies. One of the most interesting
possibilities is the violation of Lorentz symmetry \cite{ref1}. In fact, the existence of different mechanisms that bring out Lorentz-violating (LV) effects is supported in several theoretical contexts, such as loop quantum gravity \cite{ref2}, string
theory \cite{ref3}, noncommutative field theories \cite{ref3.2}, and more recently in warped brane worlds \cite{ref3.3,ref3.4} and H\v orava-Lifshitz gravity \cite{ref3.5}.

The first framework to account for LV in the SM was proposed by Colladay and Kosteleck\'y
\cite{ref4}, based on the idea of spontaneous Lorentz symmetry
breaking  in string theory \cite{refS}, known as the Standard
Model Extension (SME). The SME provides a set of gauge-invariant LV
tensor operators, compatible with the coordinate invariance
\cite{ref4L} and suitable to address the CPT and Lorentz violation
in physical systems. A number of interesting investigations have
been developed in the different sectors of the SME. The
CPT-even gauge sector was first examined by Kosteleck\'y and Mewes \cite{ref5}, with the attainment of upper bound of 1 part in $10^{37}$ (using birefrigence data). This sector was also addressed in connection with its classical solutions \cite{ref6}, consistency
aspects \cite{ref6B} and fermion/photon interactions \cite{ref6C,ref6D}. More recently, new works have proposed LV
scenarios endowed with higher dimensional operators, with new interesting
results \cite{ref6E,ref6F}. Higher dimensional operators can be
considered in terms of nonminimal interactions as well. A CPT-odd nonminimal
coupling for fermions was first regarded in Ref. \cite{ref7A}, with
some recent developments \cite{ref7B}. Very recently, an analogue
CPT-even nonminimal coupling for fermions, embracing the $K_{F}$ gauge tensor of the SME, was proposed and discussed both in relativistic and
nonrelativistic scenarios \cite{ref8}.

Another relevant SME sector much addressed in the recent years is
the gravitational one. The SME accommodates both explicit symmetry
breaking as well as spontaneous breaking. However, when one focuses on its gravitational sector, one notices that the explicit
violation is incompatible with geometrical identities like the Bianchi
identity, which suggests one should work with spontaneous
breakings to address LV within the gravitational sector \cite{ref14}. A general treatment of spontaneous local Lorentz and diffeomorphism violation for the gravitational sector of the SME was first addressed in Refs. \cite{ref15,ref16}.
In these papers, it is supposed that tensor fields acquire nonzero vacuum expectation values (VEV), breaking these symmetries spontaneously. It was then shown that the corresponding linearized effective equations can be used to study the  post-Newtonian effects in a series of gravitational systems \cite{ref17,ref18,ref19}. It is worth mentioning that a discussion for alternative ways to introduce Lorentz violation in gravity  was considered in \cite{ref19.2}.

In this paper we investigate low-energy effects of Lorentz violation
in the context of the gravitational sector of SME. More precisely, we choose a particular model in which the spontaneous Lorentz
violation comes from the dynamics of a single vector field $B_{\mu }$, coupled with the gravitational field through a term $B_{\mu}B_{\nu}R^{\mu\nu}$. This theory represents the simplest case of the well-known bumblebee models, which were first introduced by Kosteleck\'y and Samuel  in the context of string theory \citep{refS}. 
In the weak-field approximation, we determine the  modified graviton propagator and examine the effects of the Lorentz-violating background on the gravity excitations. Next, we show that the introduction of
an uncharged scalar field, coupled with the gravitational field,
leads to corrections to the classical Newtonian potential. This corrections are able at break down the radial symmetry present in standard case, revealing a spatial anisotropy due to the presence of a term proportional to $b_i b_j \hat{x}^i \hat{x}^j$. In fact, this result is corroborated by a series of post-Newtonian calculations for the puregravity sector of the minimal SME \citep{ref16, ref20, ref20.1}.
Other interesting and new term that we have found is proportional to $\nabla^{2}\frac{1}{r}\thicksim\delta^{(3)}(\vec{x})$ and it can be interpreted as a gravitational Darwin term in analogy to the usual electric Darwin term $\nabla\cdot \vec{E}$, which is generally obtained, together with spin-orbit coupling, from a nonrelativistic
limit of the Dirac equation \citep{Sakurai}. Throughout this work we shall use
the spacetime signature $(+\ -\ -\ -)$ and adopt the following definition
for the Ricci tensor: $R_{\mu \nu }=\partial _{\sigma }\Gamma _{\mu \nu
}^{\sigma }-\partial _{\nu }\Gamma _{\mu \sigma }^{\sigma }+\Gamma _{\sigma
\lambda }^{\lambda }\Gamma _{\mu \nu }^{\sigma }-\Gamma _{\sigma \nu
}^{\lambda }\Gamma _{\mu \lambda }^{\sigma },$ where $\Gamma _{\mu \nu
}^{\lambda }=\frac{1}{2}g^{\lambda \sigma }\left( \partial _{\mu }g_{\nu
\sigma }+\partial _{\nu }g_{\mu \sigma }-\partial _{\sigma }g_{\mu \nu
}\right) .$ All quantities are expressed in natural units $(\hslash
=c=\epsilon _{0}=1)$, in which the gravitational constant is $\mbox{G}%
_{N}=6.707\times 10^{-57}\mbox{eV}^{-2}$. Moreover, tensors are symmetrized
with unit weight, i.e., $A_{(\mu \nu )}=\frac{1}{2}(A_{\mu \nu }+A_{\nu \mu
})$.

The structure of the paper is as follows. Section \ref{sec:theoretical-model} is devoted to discussing the theoretical model,
introducing the general action including a LV term, and then restricting to
spontaneous LV. In Sec. \ref{sec:Weak-field-approximation-and}, we perform the weak-field approximation and calculate the
LV-corrected propagator. In Sec. \ref{sec:Modified-Newton=002019s-law}, we
introduce the coupling with a matter field and
obtain the nonrelativistic potential for two bosons interacting
gravitationally, via a scattering process. Finally, we present our final
remarks in Sec. \ref{sec:Conclusions}.

\section{The theoretical model\label{sec:theoretical-model}}

The simplest gravity model involving Lorentz-violating terms that combine
tensor fields and responsible for the spontaneous local Lorentz breaking, with
the gravitational field in $(3+1)$-dimensional Riemann
spacetime, is given by the action

\begin{equation}
S=S_{\mbox{EH}}+S_{\mbox{LV}}+S_{\mbox{matter}}.  \label{eq:total1}
\end{equation}%
The first piece in the above equation represents the usual Einstein-Hilbert
action, defined by
\begin{equation}
S_{\mbox{EH}}=\int d^{4}x\sqrt{-g}\frac{2}{\kappa ^{2}}\left( R-2\Lambda
\right) ,
\end{equation}%
where $g$ denotes the determinant of the metric field $g_{\mu \nu }$, $R$ is
the Ricci scalar, $\Lambda $ is the cosmological constant and $\kappa
^{2}=32\pi \mbox{G}_{N}$ is the gravitational coupling. Since our main goal
is to examine the effects of the Lorentz-violating on the nonrelativistic
gravitational potential, we can disregard the implications of $\Lambda $,
assuming it equal to zero hereafter.

The second piece in Eq. \eqref{eq:total1} represents the gravitational
sector for the minimal SME and contains the coefficients for Lorentz
violation, coupled to the Riemann, Ricci, and scalar curvatures, in
the following form (see, e.g., \cite{ref16}):
\begin{equation}
S_{\mbox{LV}}=\int d^{4}x\sqrt{-g}\frac{2}{\kappa ^{2}}\left( u R+s^{\mu \nu
}R_{\mu \nu }+t^{\mu \nu \alpha \beta }R_{\mu \nu \alpha \beta }\right) ,
\label{eq:LVaction}
\end{equation}%
where $u$, $s^{\mu \nu }$ and $t^{\mu \nu \alpha \beta }$ are dynamical
tensor fields with zero mass dimension and with $s^{\mu\nu}$ and $t^{\mu\nu\alpha\beta}$ having the same symmetries as the Ricci and Riemann tensors, respectively. This action is assumed to be
invariant under general coordinate transformations and the local
Lorentz violation must be achieved through a Higgs-like mechanism.

The last term on the right side of Eq. \eqref{eq:total1} takes into account the
matter-gravity couplings, which in principle should include all fields of
the standard model as well as possible interactions with coefficients $u$,  $s^{\mu\nu}$ and $t^{\mu\nu\alpha\beta}$. However, we will focus our
attention on the possible effects produced by the action \eqref{eq:LVaction},  restricting ourselves to the case where the ordinary matter only interacts
with the gravitational field. Further details about these effects in the
context of Lorentz-violation involving the matter sector of the SME can be
seen in Ref. \cite{ref20}.

Next, let us consider the particular case when $t^{\mu \nu \alpha \beta }=0$. The coefficients $u$ and $s^{\mu \nu }$ have 10 degrees of freedom (the
trace of $s^{\mu \nu }$ could be absorbed in the scalar coefficient
$u$) that may be described by an effective field theory involving a single
vector field $B_{\mu }$, whose dynamics is determined by the following
action:
\begin{equation}
S_{\mbox{B}}=\int d^{4}x\sqrt{-g}\left[ -\frac{1}{4}B_{\mu \nu }B^{\mu \nu
}+\sigma B^{\mu }B^{\nu }R_{\mu \nu }-V(B_{\mu }B^{\mu }\mp b^{2})\right] ,
\label{eq:bumblebee1}
\end{equation}%
where $B_{\mu \nu }=\partial _{\mu }B_{\nu }-\partial _{\nu }B_{\mu }$ , $\sigma$ is a dimensionless coupling constant and $b^{2}$ is a positive
constant that sets the VEV for $B_{\mu}$. The potential $V(x)$
triggers the spontaneous breakdown of both Lorentz and diffeomorphism
symmetries, such that its minimum occurs at $g^{\mu \nu }B_{\mu }B_{\nu }\pm
b^{2}=0$, i.e., when $B_{\mu }$ and $g_{\mu \nu }$ acquire nonzero vacuum
expectation values. This theory is a particular case of the so-called
bumblebee models and were initially evaluated in the context of string
theory \cite{refS}. Furthermore, we note that for $\sigma =0$ the action for
the bumblebee field becomes $U(1)$ gauge invariant and the potential $V$
also spontaneously breaks this symmetry.

The correspondence between the action \eqref{eq:LVaction} and the bumblebee
model \eqref{eq:bumblebee1} is obtained through the relations \cite{ref16},
\begin{eqnarray}
u & = & \frac{1}{4}\xi B^{\alpha}B_{\alpha},  \notag \\
s^{\mu\nu} & = & \xi B^{\mu}B^{\nu}-\frac{1}{4}\xi
g^{\mu\nu}B^{\alpha}B_{\alpha},  \notag \\
t^{\mu\nu\alpha\beta} & = & 0,  \label{eq:LIVBumblebeeRelation1}
\end{eqnarray}
where for convenience we write $\sigma=(2\xi/\kappa^{2})$, so that the mass dimension of the bumblebee field and the coupling constant are, respectively: $\left[B^{\mu}\right]=1$, $\left[\xi\right]=-2$.

\section{Weak-field approximation and the graviton propagator\label%
{sec:Weak-field-approximation-and}}

To investigate the effects of gravity-bumblebee coupling on the graviton
dynamics, we split the dynamical fields into the vacuum expectation values
and the quantum fluctuations,
\begin{eqnarray}
g_{\mu\nu} & = & \eta_{\mu\nu}+\kappa h_{\mu\nu},  \notag \\
B_{\mu} & = & b_{\mu}+\tilde{B}_{\mu},  \notag \\
B^{\mu} & = & b^{\mu}+\tilde{B}^{\mu}-\kappa b_{\nu}h^{\mu\nu},
\label{eq:Expansion1}
\end{eqnarray}
where $h_{\mu\nu}$ and $\tilde{B}_{\mu}$ represent small perturbations
around the Minkowski background and a constant vacuum value $b_{\mu}$,
respectively. The vector $b_{\mu}$ is the local Lorentz violation
coefficient associated to the bumblebee field.

Varying the action \eqref{eq:bumblebee1} with respect to $B_{\mu }$,
we obtain the equation of motion for the bumblebee field,
\begin{equation}
\frac{1}{\sqrt{-g}}\partial ^{\mu }\left\{ \sqrt{-g}B_{\mu \nu }\right\}
-2V^{\prime }B_{\nu }+2\sigma B^{\mu }R_{\mu \nu }=0,
\label{eq:Motion1Bumblebee}
\end{equation}
where the prime on $V$ means differentiation with respect to the
argument.

Following the ideas described in Ref \cite{ref16}, we may employ the
expansions defined in Eq. \eqref{eq:Expansion1}, and assume for $V(x)$ the
smooth quadratic form
\begin{equation}
V=\frac{\lambda }{2}\left( B^{\mu }B_{\mu }\mp b^{2}\right) ^{2},
\end{equation}%
so that the linearized version of the equation of motion
\eqref{eq:Motion1Bumblebee} can be written as
\begin{equation}
\left( \square \eta _{\mu \nu }-\partial _{\mu }\partial _{\nu }-4\lambda
b_{\mu }b_{\nu }\right) \tilde{B}^{\mu }=-2\lambda \kappa b_{\nu }b_{\alpha
}b_{\beta }h^{\alpha \beta }-2\sigma b^{\alpha }R_{\alpha \nu },
\label{eq:motionBumblebeeLinear}
\end{equation}%
with $\square \equiv \partial ^{2}$. In this expression, $R_{\mu \nu }$
shall be understood as being in its linearized form. Also, for simplicity $b_{\mu}$ is adopted as a timelike vector, such that $b^{\mu }b_{\mu }=+b^{2}
$. Applying the Green's function method, the solution to Eq.
\eqref{eq:motionBumblebeeLinear} is straightforward, leading in momentum
space to the the following expression:
\begin{equation}
\tilde{B}^{\mu }=\frac{\kappa p^{\mu }b_{\alpha }b_{\beta }h^{\alpha \beta}
}{2b\cdot p}+\frac{2\sigma b_{\alpha }R^{\alpha \mu }}{p^{2}}-\frac{2\sigma
p^{\mu }b_{\alpha }b_{\beta }R^{\alpha \beta }}{p^{2}b\cdot p}+\frac{\sigma
p^{\mu }R}{4\lambda b\cdot p}-\frac{\sigma b^{\mu }R}{p^{2}}+\frac{\sigma
p^{\mu }b^{2}R}{p^{2}b\cdot p},
\end{equation}
with $b\cdot p=b_{\mu }p^{\mu }$, $p^{2}=p\cdot p=p_{\mu }p^{\mu}$.

By substituting this solution into the action \eqref{eq:LVaction},
with the help of the relations defined by Eqs.
\eqref{eq:LIVBumblebeeRelation1} and \eqref{eq:Expansion1} in a suitable
order, we are able to determine the modifications yielded by the nonzero
vacuum expectation value $b_{\mu }$ on the kinetic terms of the graviton
field. Therefore, it is necessary to expand the bumblebee-graviton
interaction $\mathcal{L}_{\mbox{LV}}$ up to second order in $h_{\mu \nu }$,
as follows,
\begin{eqnarray}
\mathcal{L}_{\mbox{LV}} &=&\sigma \sqrt{-g}B^{\mu }B^{\nu }R_{\mu \nu }
\notag \\
&=&\sigma \left[ b_{\mu }b_{\nu }R^{\mu \nu }(h^{2})+2b_{\mu }\tilde{B}_{\nu
}R^{\mu \nu }(h)+\frac{1}{2}\kappa h_{\ \alpha }^{\alpha }b_{\mu }b_{\nu
}R^{\mu \nu }(h)\right] +\mathcal{O}(h^{3}),
\end{eqnarray}%
where the order in $h_{\mu \nu }$ at the Ricci tensors is explicitly
indicated. Replacing $\tilde{B}^{\mu }$ and grouping the terms
conveniently, we obtain
\begin{eqnarray}
\mathcal{L}_{\mbox{LV}} &=&\xi \left[ p^{2}b_{\mu }b_{\nu }h^{\mu \nu }h_{\
\alpha }^{\alpha }+\frac{1}{2}\left( b\cdot p\right) ^{2}\left( h_{\ \alpha
}^{\alpha }\right) ^{2}\right.   \notag \\
&-&\left. \frac{1}{2}\left( b\cdot p\right) ^{2}h^{\mu \nu }h_{\mu \nu
}+p^{2}b_{\mu }b_{\nu }h^{\mu \alpha }h_{\ \alpha }^{\nu }-\left( b_{\mu
}b_{\nu }p_{\alpha }p_{\beta }+b_{(\mu }p_{\nu )}b_{(\alpha }p_{\beta
)}\right) h^{\mu \nu }h^{\alpha \beta }\right]   \notag \\
&+&\frac{4\xi ^{2}}{\kappa ^{2}}\left[ \left( -2p^{2}b_{\mu }b_{\nu
}-2b^{2}p_{\mu }p_{\nu }+4b\cdot pb_{(\mu }p_{\nu )}-\frac{p^{2}p_{\mu
}p_{\nu }}{4\lambda }\right) h^{\mu \nu }h_{\ \alpha }^{\alpha }\right.
\notag \\
&+&\left( 2b_{\mu }b_{\nu }p_{\alpha }p_{\beta }-b_{(\mu }p_{\nu
)}b_{(\alpha }p_{\beta )}+\frac{b^{2}p_{\mu }p_{\nu }p_{\alpha }p_{\beta }}{%
p^{2}}-\frac{2b\cdot pp_{\mu }p_{\nu }b_{(\alpha }p_{\beta )}}{p^{2}}+\frac{%
p_{\mu }p_{\nu }p_{\alpha }p_{\beta }}{4\lambda }\right) h^{\mu \nu
}h^{\alpha \beta }  \notag \\
&+&\left. \left( b^{2}p^{2}-\left( b\cdot p\right) ^{2}+\frac{p^{4}}{%
4\lambda }\right) \left( h_{\ \alpha }^{\alpha }\right) ^{2}+\left(
p^{2}b_{\mu }b_{\nu }-2b\cdot pb_{(\mu }p_{\nu )}+\frac{\left( b\cdot
p\right) ^{2}p_{\mu }p_{\nu }}{p^{2}}\right) h^{\mu \lambda }h_{\ \lambda
}^{\nu }\right] +\mathcal{O}(h^{3}),  \notag \\
&&
\label{LagrangianLV}\end{eqnarray}
with $\sigma =(2\xi /\kappa ^{2})$, as previously defined. It should be noted that the first-order terms in the gravity-bumblebee coupling constant $\xi$ are all quadratic in the background  $b^{\mu}$, 
but in second-order $\mathcal{O}(\xi^{2})$, there are contributions which are background independent, and that come from the $\lambda$ term in the bumblebee fluctuation $\tilde{B}^{\mu}$. These contributions introduce higher derivatives corrections  $(\partial ^ {4})$ on the kinetic term of the graviton field. As we shall see in the next section, these two kinds of modifications will induce different corrections on the gravitational potential.

The Lorentz-violating Lagrangian \eqref{LagrangianLV} can be rewritten to position space and combined with the
expanded Einstein-Hilbert Lagrangian,
\begin{equation}
\mathcal{L}_{\mbox{EH}}=\partial h^{\mu \nu }\partial _{\alpha }h_{\nu
}^{\alpha }-\partial _{\mu }h^{\mu \nu }\partial _{\nu }h+\frac{1}{2}%
\partial _{\mu }h\partial ^{\mu }h-\frac{1}{2}\partial _{\alpha }h^{\mu \nu
}\partial ^{\alpha }h_{\mu \nu }+\mathcal{O}(h^{3}),
\end{equation}
with $h\equiv h_{\ \lambda }^{\lambda }$. We add one convenient
gauge fixing term,
\begin{equation}
\mathcal{L}_{\mbox{gf}}=-(\partial _{\mu }h^{\mu \nu }-\frac{1}{2}\partial
^{\nu }h)^{2},  \label{eq:GaugeFixing}
\end{equation}
to yield the effective Lagrangian, which we need to consider in order to
obtain the modified graviton propagator. Then, in the $\mathcal{L}_{\mbox{EH}}+\mathcal{L}_{\mbox{gf}}+\mathcal{L}_{\mbox{LV }}$, the kinetic
term for the graviton field becomes
\begin{equation}
\mathcal{L}_{\mbox{kin}}=-\frac{1}{2}h^{\mu \nu }\hat{\mathcal{O}}_{\mu \nu
,\alpha \beta }h^{\alpha \beta },
\end{equation}
where the operator $\hat{\mathcal{O}}_{\mu \nu ,\alpha \beta }$ is separated
in two pieces
\begin{equation}
\hat{\mathcal{O}}_{\mu \nu ,\alpha \beta }=\hat{\mathcal{K}}_{\mu \nu
,\alpha \beta }+\hat{\mathcal{V}}_{\mu \nu ,\alpha \beta },
\label{eq:operator1}
\end{equation}
such that $\hat{\mathcal{K}}_{\mu \nu ,\alpha \beta }$ is the usual
quadratic form,
\begin{equation}
\hat{\mathcal{K}}_{\mu \nu ,\alpha \beta }=\frac{1}{2}\left( \eta _{\mu
\alpha }\eta _{\nu \beta }+\eta _{\mu \beta }\eta _{\nu \alpha }-\eta _{\mu
\nu }\eta _{\alpha \beta }\right) (-\partial ^{2}),
\end{equation}
while $\hat{\mathcal{V}}_{\mu \nu ,\alpha \beta }$ encloses the terms that
contain the Lorentz-violating Lagrangian $\mathcal{L}_{\mbox{LV}}$.

The graviton propagator is defined by
\begin{equation}
\left\langle 0\left|T\left[h_{\mu\nu}(x)h_{\alpha\beta}(y)\right]
\right|0\right\rangle =D_{\mu\nu,\alpha\beta}(x-y),  \label{eq:prop1}
\end{equation}
where $D_{\mu\nu,\alpha\beta}$ is the operator that satisfies the Green
\textquoteright{}s equation, given as
\begin{equation}
\hat{\mathcal{O}}_{\ \
\lambda\sigma}^{\mu\nu,}D^{\lambda\sigma,\alpha\beta}(x-y)=i\mathcal{I}
^{\mu\nu,\alpha\beta}\delta^{4}(x-y),
\end{equation}
with $\mathcal{I}^{\mu\nu,\alpha\beta}=\frac{1}{2}\left(\eta^{\mu\alpha}
\eta^{\nu\beta}+\eta^{\mu\beta}\eta^{\nu\alpha}\right)$. Thus, the exact
graviton propagator is evaluated by inverting \eqref{eq:operator1}, finding
a closed operator algebra composed by a set of appropriated projectors.
It is known that the bumblebee model under study has Nambu-Goldstone and massive propagating modes \cite{ref15}. The implications of these modes on the graviton propagator, concerning the stability, causality and unitarity of this theory are important issues that have not been investigated so far. However, the full calculation of the graviton propagator on the presence of Lorentz violation is not the main purpose of the present work and will be addressed in an upcoming work. Thus motivated by the fact that the
magnitude of $b_{\mu}$ should be small  as well as the coupling constant $\xi$, we make use the conventional graviton
propagator in the gauge given by Eq. \eqref{eq:GaugeFixing} and treat the
Lorentz-violating term in Eq. \eqref{eq:operator1} as a perturbative
insertion \cite{ref22}. This is accomplished by means of the following matricial
identity:
\begin{equation}
\frac{1}{A+B}=\frac{1}{A}-\frac{1}{A}B\frac{1}{A+B}=\frac{1}{A}-\frac{1}{A}B
\frac{1}{A}+\frac{1}{A}B\frac{1}{A}B\frac{1}{A+B}=\cdots.
\end{equation}

The operator $\hat{\mathcal{K}}$ can easily be inverted and the conventional
graviton propagator is then written in the momentum space as
\begin{equation}
D_{0}^{\mu\nu,\alpha\beta}(q)=\frac{i}{2}\frac{\eta^{\mu\alpha}\eta^{\nu
\beta}+\eta^{\mu\beta}\eta^{\nu\alpha}-\eta^{\mu\nu}\eta^{\alpha\beta}}{
q^{2}+i\epsilon}.
\end{equation}

After lengthy contraction operations of indices, we are ready to give the
explicit form of $D^{\mu\nu,\alpha\beta}=D_{0}^{\mu\nu,\alpha\beta}+D_{\mbox{LV}}^{\mu\nu,\alpha\beta}$ up to second order in $b_{\mu}$, which
reads as
\begin{eqnarray}
\left(D_{\mbox{LV}}^{\mu\nu,\alpha\beta}\right)_{\xi} & = & i\xi\left[b^{2}\left(\frac{g^{\alpha\beta}g^{\mu\nu}}{q^{2}}+\frac{q^{\alpha}q^{\beta}g^{\mu\nu}}{q^{4}}\right)\right.  \notag \\
& + & \frac{\left(b\cdot
q\right)^{2}\left(g^{\alpha\beta}g^{\mu\nu}-g^{\alpha\nu}g^{\beta\mu}-g^{\alpha\mu}g^{\beta\nu}\right)}{2q^{4}}  \notag \\
& + & \frac{b\cdot
q\left(b^{\beta}q^{\alpha}g^{\mu\nu}+b^{\alpha}q^{\beta}g^{\mu\nu}+b^{\nu}q^{\mu}g^{\alpha\beta}+b^{\mu}q^{\nu}g^{\alpha\beta}\right)}{2q^{4}}
\notag \\
& + & \left(\frac{b^{\alpha}b^{\mu}g^{\beta\nu}+b^{\beta}b^{\mu}g^{\alpha\nu}+b^{\alpha}b^{\nu}g^{\beta\mu}+b^{\beta}b^{\nu}g^{\alpha\mu}-2b^{\alpha}b^{\beta}g^{\mu\nu}-4b^{\mu}b^{\nu}g^{\alpha\beta}}{2q^{2}}\right.
\notag \\
& - & \left.\left.\frac{4b^{\mu}b^{\nu}q^{\alpha}q^{\beta}+b^{\beta}b^{\nu}q^{\alpha}q^{\mu}+b^{\alpha}b^{\nu}q^{\beta}q^{\mu}+b^{\beta}b^{\mu}q^{\alpha}q^{\nu}+b^{\alpha}b^{\mu}q^{\beta}q^{\nu}}{2q^{4}}\right)\right],
\end{eqnarray}

\begin{eqnarray}
\left(D_{\mbox{LV}}^{\mu\nu,\alpha\beta}\right)_{\xi^{2}} & = & \frac{
i\xi^{2}}{\kappa^{2}}\left[b^{2}\left(\frac{12q^{\mu}q^{\nu}g^{\alpha
\beta}-12q^{\alpha}q^{\beta}g^{\mu\nu}}{q^{4}}+\frac{8q^{\alpha}q^{\beta}q^{
\mu}q^{\nu}}{q^{6}}\right)+\frac{g^{\alpha\beta}g^{\mu\nu}}{2\lambda}\right.
\notag \\
& + & \frac{2\left(b\cdot
q\right)^{2}\left(q^{\alpha}q^{\mu}g^{\beta\nu}+q^{\beta}q^{\mu}g^{\alpha
\nu}+q^{\alpha}q^{\nu}g^{\beta\mu}+q^{\beta}q^{\nu}g^{\alpha\mu}+2q^{\mu}q^{
\nu}g^{\alpha\beta}-2q^{\alpha}q^{\beta}g^{\mu\nu}\right)}{q^{6}}  \notag \\
& + & b\cdot q\left\{ \frac{10\left(b^{\beta}q^{\alpha}g^{\mu\nu}+b^{
\alpha}q^{\beta}g^{\mu\nu}-b^{\nu}q^{\mu}g^{\alpha\beta}-b^{\mu}q^{\nu}g^{
\alpha\beta}\right)}{q^{4}}+\frac{8\left(b^{\beta}q^{\alpha}q^{\mu}q^{
\nu}+b^{\alpha}q^{\beta}q^{\mu}q^{\nu}\right)}{q^{6}}\right.  \notag \\
& - & \left.\frac{4\left(b^{\mu}q^{\alpha}g^{\beta\nu}-b^{\mu}q^{\beta}g^{
\alpha\nu}-b^{\nu}q^{\alpha}g^{\beta\mu}-b^{\nu}q^{\beta}g^{\alpha\mu}\right)
}{q^{4}}\right\} -\frac{q^{\alpha}q^{\beta}g^{\mu\nu}}{q^{2}\lambda}+\frac{
3q^{\mu}q^{\nu}g^{\alpha\beta}}{q^{2}\lambda}  \notag \\
& + & \left\{ \frac{2\left(b^{\alpha}b^{\mu}g^{\beta\nu}+b^{\beta}b^{\mu}g^{
\alpha\nu}+b^{\alpha}b^{\nu}g^{\beta\mu}+b^{\beta}b^{\nu}g^{\alpha\mu}-2b^{
\alpha}b^{\beta}g^{\mu\nu}+2b^{\mu}b^{\nu}g^{\alpha\beta}\right)}{q^{2}}
\right.  \notag \\
& + & \left.\left.\frac{2\left(8b^{\mu}b^{\nu}q^{\alpha}q^{\beta}-b^{
\beta}b^{\nu}q^{\alpha}q^{\mu}-b^{\alpha}b^{\nu}q^{\beta}q^{\mu}-b^{
\beta}b^{\mu}q^{\alpha}q^{\nu}-b^{\alpha}b^{\mu}q^{\beta}q^{\nu}\right)+
\frac{2q^{\alpha}q^{\beta}q^{\mu}q^{\nu}}{\lambda}}{q^{4}}\right\} \right],
\end{eqnarray}
where $\left(D_{\mbox{LV}}^{\mu\nu,\alpha\beta}\right)_{\xi}$ and $\left(D_{
\mbox{LV}}^{\mu\nu,\alpha\beta}\right)_{\xi^{2}}$are contributions to $D_{
\mbox{LV}}^{\mu\nu,\alpha\beta}$ proportional to $\xi$ and $\xi^{2}$,
respectively.

Some comments about these results are worthwhile. Taking into account the
expression \eqref{eq:LIVBumblebeeRelation1}, the products $b^{2}$, $(b\cdot
q)^{2}$ and $(b\cdot q)b^{\mu }$ are first order terms in the
Lorentz-violating coefficients $u$ and $s^{\mu \nu }$. Thus, we note that
the correction $\left( D_{\mbox{LV}}^{\mu \nu ,\alpha \beta }\right) _{\xi }$
involves only terms to first order in $u$ and $s^{\mu \nu }$, and do not depend
on the particular form of the bumblebee
potential $V(x)$. In second order at $\xi$, there are terms that are not associated with the vector $b_{\mu }$
(they are proportional to $\lambda ^{-1}$) and depend only on the coupling of
that potential. In addition, the corrections to graviton propagator have
poles in $q^{2}=0$, showed that in this approximation the theory is free of
ghosts and tachyons. Nevertheless, the expression for $\left( D_{\mbox{LV}}^{\mu \nu ,\alpha \beta }\right) _{\xi^2}$ also possess a nonpole term $g^{\alpha\beta}g^{\mu\nu}/2\lambda$ that may be related to the propagation of massive bumblebee mode in the graviton propagator. In fact, the analytic contributions which are generated by massive particles in the Feynman diagrams can be expanded in a Taylor serie as  $1/(q^2-m^2)=-1/m^2(1-q^2/m^2+\cdots)$  \cite{ref23, Donoghue}.  Thus, only when we evaluate the  tree-level graviton propagator in an exact tensor form, we will be able to answer if there are nonphysical modes induced by the higher derivative terms and the Lorentz-violating term. Any way, the treatment of Lagrangian \eqref{LagrangianLV} as a perturbative insertion can be performed, and it represents a reasonable approximation. Finally, it is still important to mention that this
propagator is symmetric under an indices permutation $(\mu \leftrightarrow
\nu )$ and $(\alpha \leftrightarrow \beta )$, as it really must be. We should draw attention for other evaluations concerning the
graviton propagator \cite{refH}.

\section{Modified Newton\textquoteright{s} law of gravitation \label{sec:Modified-Newton=002019s-law}}

In this section, we study the effects of the spontaneous Lorentz violation
when we consider the tree-level modified propagator as determined previously. One
of the simplest examples that we can choose to evaluate such effects,
consists in the gravitational interaction of two distinguishable heavy
particles described in the nonrelativistic limit by the Newtonian potential.
Thus, our main goal here is to determine the scattering amplitude of two massive
bosons particles of spin-zero by one-graviton exchange. Once calculated the
matrix amplitude in leading order, we can take the nonrelativistic limit and
compare it with the Born approximation to determine the potential modified
by the nonzero vacuum expectation value $b_{\mu}$.

Consider the following action for a real scalar field in curved spacetime,
\begin{equation}
S_{\mbox{matter}}=\int d^{4}x\sqrt{-g}\left[\frac{1}{2}g^{\mu\nu}\partial_{\mu}\phi\partial_{\nu}\phi-\frac{1}{2}m^{2}\phi^{2}\right],
\end{equation}
which can be expanded in the weak field approximation up to first order in $h$. We are then left with the following Lagrangian:
\begin{eqnarray}
\mathcal{L}_{\mbox{matter}} & \thickapprox & \frac{1}{2}\partial_{\mu}\phi%
\partial^{\mu}\phi-\frac{1}{2}m^{2}\phi^{2}  \notag \\
& & -\frac{1}{2}\kappa h^{\mu\nu}\left[\partial_{\mu}\phi\partial_{\nu}\phi-%
\frac{1}{2}\eta_{\mu\nu}\left(\partial_{\alpha}\phi\partial^{\alpha}%
\phi-m^{2}\phi^{2}\right)\right].
\end{eqnarray}

\begin{figure}[tbp]
\begin{centering}
\includegraphics[scale=1]{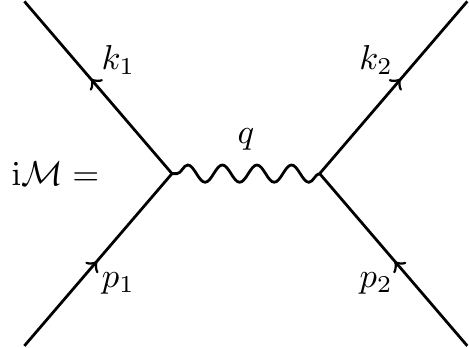}
\par\end{centering}
\caption{The tree-level diagram of two scalar particles interacting via the
exchange of a graviton.}
\label{fig:1}
\end{figure}

Now, let us consider the scattering process involving two scalar particles of
mass $m_{1}$ and $m_{2}$. The only Feynman diagram that contributes to this
process, in lowest order, is drawn in Fig. \ref{fig:1}, and its analytical
expression can be written as
\begin{equation}
i\mathcal{M}=(-i\kappa)^{2}V^{\mu\nu}(p_{1},-k_{1},m_{1})D_{\mu\nu,\alpha%
\beta}(q)V^{\alpha\beta}(p_{2},-k_{2},m_{2}),  \label{eq:matrix1}
\end{equation}
where $q=p_{2}-k_{2}=-(p_{1}-k_{1})$ is the momentum transfer and the vertex
$V^{\mu\nu}(p,k,m)$ corresponds to the expression
\begin{equation}
V^{\mu\nu}(p,k,m)=-\frac{1}{2}\left[p^{\mu}k^{\nu}+p^{\nu}k^{\mu}-\eta^{\mu%
\nu}\left(p\cdot k+m^{2}\right)\right].  \label{eq:vertex}
\end{equation}

Substituting the expressions defined in \eqref{eq:prop1} and %
\eqref{eq:vertex} into the scattering amplitude \eqref{eq:matrix1}, we
arrive at the sum of the two pieces:
\begin{equation}
i\mathcal{M}=i\mathcal{M}_{0}+i\mathcal{M}_{\mbox{LV}}\label{matrixLIV},
\end{equation}
such that the first term is just the conventional amplitude given by \cite%
{ref23}
\begin{eqnarray}
i\mathcal{M}_{0} & = & -\frac{i\kappa^{2}}{8q^{2}}\left[4\left\{ k_{1}\cdot
p_{1}\left(m_{2}^{2}-k_{2}\cdot p_{2}\right)+k_{1}\cdot p_{2}k_{2}\cdot
p_{1}+k_{1}\cdot k_{2}p_{1}\cdot p_{2}\right\} \right.  \notag \\
& & \left.-2m_{1}^{2}\left\{ 4\left(m_{2}^{2}-k_{2}\cdot
p_{2}\right)+2k_{2}\cdot p_{2}\right\} \right],
\end{eqnarray}
which is modified by $i\mathcal{M}_{\mbox{LV}}$, consisting of a large expression involving the possible contractions
of $b^{\mu}$ with the four-momenta of the incoming and outgoing scalar field and also with the virtual graviton momentum.

To access the nonrelativistic limit, we take the approximation (also called static limit) $p_{1,2}=(m_{1,2},0),$ $k_{1,2}=(m_{1,2},0),$ and $q=(0,\vec{q})$. In this way, the scalar products involving $b^{\mu}$ can be written as follows: $b\cdot p_{1,2}=b\cdot k_{1,2}=b_{0}m_{1,2}$ and $b\cdot q=-(\vec{b}\cdot\vec{q})$, so that $b^{\mu}=(b^{0},\vec{b})$ is the
constant background in an asymptotically inertial frame.

Inserting these expressions into the matrix amplitude \eqref{matrixLIV} and collecting the remaining terms, we get the simplified result
\begin{equation}
i\mathcal{M}_{\mbox{NR}} = \frac{i\kappa^{2}m_{1}^{2}m_{2}^{2}}{2\vec{q}^{2}}
- \frac{i\xi \vec{b}^{2}\kappa^{2}m_{1}^{2}m_{2}^{2}}{\vec{q}^{2}}+\frac{i\xi \left(\vec{b}\cdot \vec{q}\right)^{2}\kappa^{2}m_{1}^{2}m_{2}^{2}}{2 \vec{q}^{4}}+\frac{
8i\xi^{2}b_{0}^{2}m_{1}^{2}m_{2}^{2}}{\vec{q}^{2}}-\frac{%
i\xi^{2}m_{1}^{2}m_{2}^{2}}{2\lambda},  \label{eq:NonrelatMatrix}
\end{equation}
where the first term gives the well-known tree-level result, whose Fourier transform yields the standard Newtonian potential, while the other terms represent the matrix elements arising from the spontaneous Lorentz breaking. The second and fourth terms only yield an unobservable scaling, since they can always be absorbed into the definition of the coupling constant.  However, the third and last terms contribute to the matrix element with a nontrivial physical  and will be discussed below.

To make the connection to the Newtonian gravitational potential, we follow
Ref. \cite{ref24}, and define the potential Fourier transformed in the
nonrelativistic limit by
\begin{eqnarray}
\left\langle \mbox{f}\left|i\mbox{T}\right|\mbox{i}\right\rangle & \equiv &
(2\pi)^{4}\delta^{4}(p-k)i\mathcal{M}(p_{1},p_{2}\rightarrow k_{1},k_{2})
\notag \\
& \thickapprox & -(2\pi)\delta(E_{p}-E_{k})i\tilde{V}(\vec{q}),
\end{eqnarray} so that the potential in coordinate space corresponds to
\begin{equation}
V(\vec{x}) =  \frac{1}{2m_{1}}\frac{1}{2m_{2}}\int\frac{d^{3}q}{(2\pi)^{3}
}e^{i\vec{q}\cdot\vec{x}}\tilde{V}(\vec{q})\label{potCoordinate}.
\end{equation}

\begin{figure}[tbp]
\begin{centering}
\includegraphics[scale=1]{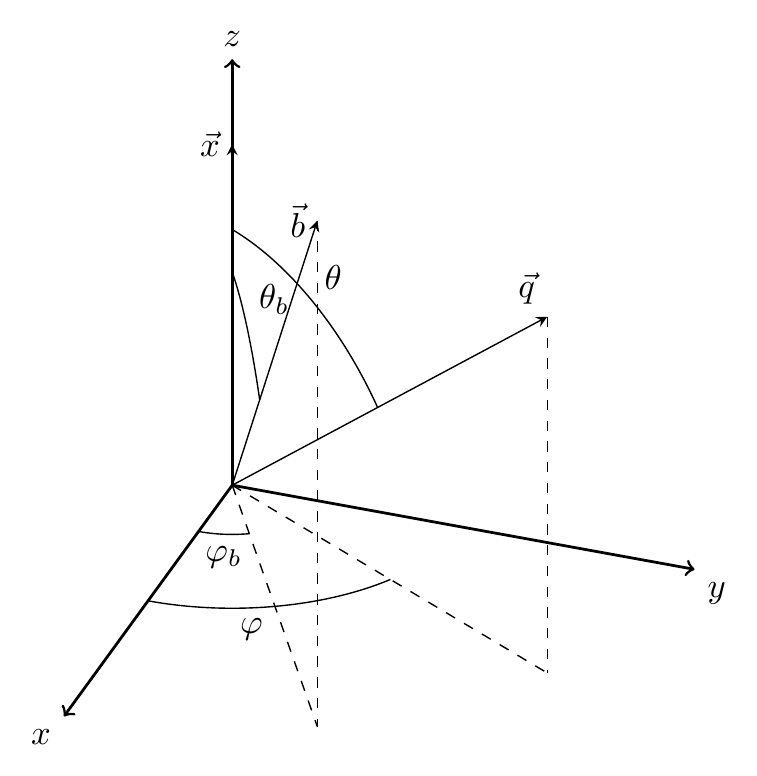}
\par\end{centering}
\caption{Definitions for the vectors and angles of interest in a standard Cartesian coordinates system.}
\label{fig:2}
\end{figure}

In order to solve Eq. \eqref{potCoordinate}, we will assume that the two point masses $m_1$ and $m_2$ are located by the coordinate vectors $\vec{x}_1$ and $\vec{x}_2$ with $\vec{x}=\vec{x}_1-\vec{x}_2$, in an inertial Cartesian coordinate system (for example, taking $m_{1} = \mbox{Sun mass}$, then this coincides with the canonical Sun-centered frame). Considering the vectors $\vec{x}$, $\vec{q}$ and $\vec{b}$ as depicted in Fig. \ref{fig:2}, we can define the following angular relations: $\cos\theta=\vec{q}\cdot\vec{x}/q r$, $\cos\theta_{b}=\vec{b}\cdot\vec{x}/b r$, $\cos\Psi=\vec{b}\cdot\vec{q}/bq$ with $\cos\Psi=\sin\theta\sin\theta_{b}\cos(\varphi-\varphi_b)+\cos\theta\cos\theta_b$,  $q=|\vec{q}|$, $r=|\vec{x}|$ and $b=|\vec{b}|$. Thus, the background vector, $\vec{b}$ , sets up a fixed direction in space, where $\theta_{b}$ and $\varphi_{b}$ are the (fixed) angles that indicate the directional dependence of the potential $V(\vec{x})$ in relation to the background direction. These expressions allow the evaluation of the angular integration on the $\Psi$ variable enclosed in Eq. \eqref{potCoordinate},
\begin{equation}
\int_{0}^{\infty}dq \int_{0}^{\pi}d\theta\sin\theta\int_{0}^{2\pi}d\varphi e^{iqr\cos\theta}\cos^{2}\Psi=\frac{\pi^2\sin^{2}\theta_{b}}{r}.\label{IntAngular}
\end{equation}

Taking into account these preliminary results, we can now calculate the momentum integral on the $q$-variable, obtaining the following Newtonian potential: 
\begin{eqnarray}
V(\vec{x}) & = & -\frac{\mbox{G}_{N} m_{1} m_{2}}{r}\left[1-
\frac{3}{2}\xi\vec{b}^{2}-\frac{1}{2}\xi\left(\vec{b}\cdot\hat{x}\right)^{2}\right]-\mbox{G}_{N}
m_{1}m_{2}\left[\frac{\xi^{2}b_{0}^{2}}{2\pi\mbox{G}_{N}}\frac{1}{r}-\frac{\xi^{2}}
{8\lambda\mbox{G}_{N}}\delta^{3}(\vec{x})\right],\nonumber\\
 \label{potCorrec}
\end{eqnarray} where $\hat{x}=\vec{x}/|\vec{x}|$. 
We note that to first-order corrections in $\xi$, the Newton\textquoteright{s}
potential remains exhibiting the standard behavior, inversely proportional to the separation distance between the two point masses. Besides, it contains an unusual directional dependence in terms of the angle $\theta_{b}$ relative to the scalar product between the background $\vec{b}$ and the unit vector $\hat{x}$ (aligned along the direction from $m_{1}$ to $m_{2}$). The attractiveness  of these corrections depend on the sign of the coupling constant $\xi$: it will be attractive for $\xi < 0$ or repulsive
for $\xi > 0$. It is worth noting that, at leading order in $\xi$, our results  are in complete agreement with those obtained in Refs \cite{ref16, ref20} from a direct calculation of the post-Newtonian metric for the pure-gravity sector of the minimal SME. In fact, if  we set $\bar{u}= \xi b^{\alpha}b_{\alpha}= 0$ (such that $\vec{b}^2=b_{0}^{2}$), but with $\xi$ replaced by $-\xi$, then the conditions \eqref{eq:LIVBumblebeeRelation1} ensure that we can rewrite the potential $V(\vec{x})$ as: 
\begin{equation}
V(\vec{x})=-\frac{\mbox{G}_{N} m_{1} m_{2}}{r}\left[1+
\frac{3}{2}\bar{s}^{00}+\frac{1}{2}\bar{s}^{ij}\hat{x}^{i}\hat{x}^{j}\right]+\cdots,
\end{equation}
which in turn has the same form as that achieved from the equation (35) of Ref. \cite{ref16}. 

In the literature \cite{ref28} there are several discussions on sensitive tests of gravity, able to establish experimental bounds on the Lorentz-violating coefficients.  A  well known example of this kind of test involves accurate measurement of the deflection angle in which a light ray is deflected by a massive body \cite{ref29}. A detailed investigation searching for deviations from the standard GR result due to the Lorentz violation has been recently performed in Ref. \cite{ref20.1}, where the deflection angle was derived directly from the post-Newtonian metric for the minimal SME. In this paper is reported that the coefficient $\bar{s}_{ij}$ is currently constrained at the $10^{-5}-10^{-6}$. These results can be used  to set up bounds on the backgound $\vec{b}$, responsible for the anisotropic effects present in our calculation for the gravitational potential,  and consequently we can assume a similar restriction on the $|\xi|b_{i}b_{j}$.

The last term in Eq. \eqref{potCorrec} provides a nontrivial contribution, involving a Dirac delta function. This short-ranged correction looks like a gravitational Darwin term and it is induced by higher derivative terms of order $\partial^{4}$ contained in the Lagrangian \eqref{LagrangianLV} at $\mathcal{O}(\xi^{2})$. Indeed, an analogue correction is observed when we add higher-order terms in the curvature  to the pure-gravity Lagrangian \cite{Donoghue}. To gain insight into the nature of this term, let us consider a simplified model defined by the Lagrangian
\begin{equation}
\mathcal{L}_{\mbox{grav}}=\sqrt{-g}\left[\frac{2}{\kappa^{2}}R+\alpha R^{2}\right],
\end{equation}where the $\alpha$ parameter is a dimensionless constant which must be determined by experiments. In the low-energy limit the effect of $R^{2}$ is add to the Newtonian potential a Yukawa potential of the form 
\begin{equation}
V(\vec{x})=-\mbox{G}_{N} m_{1} m_{2}\left[\frac{1}{r}-\frac{e^{-r/\sqrt{\kappa^{2}\alpha}}}{r}
\right].
\end{equation}

Experimental constraints on the parameter $\alpha$ are very poor and exploit deviations from the inverse square law, bounding $\alpha<10^{60}$ \cite{Accioly}. For $\sqrt{\kappa^{2}\alpha}$ small, which in practice  should be considered for a perturbation in an effective field theory, we can replace the Yukawa potential by a representation of a delta function
\begin{equation*}
\frac{e^{-r/\sqrt{\kappa^{2}\alpha}}}{r}\rightarrow 4\pi\kappa^{2}\alpha\delta^{3}(\vec{x}),
\end{equation*}which yields the following low-energy potential: 
\begin{equation}
V(\vec{x})=-\mbox{G}_{N} m_{1} m_{2}\left[\frac{1}{r}-
128\pi^{2}\mbox{G}_{N}\alpha\delta^{3}(\vec{x})\right].
\end{equation}

So, the higher-order term $R^{2}$ gives rise to a very small and short-ranged modification to the Newtonian potential and has the same form as that obtained to the last term in Eq. \eqref{potCorrec}. In recent gravitational experiments, it is found that the Newtonian gravitational interaction, seems to be maintained up to $\thicksim 0.13-0.16$ mm \cite{Yang}.  A detailed analysis of this experiment on the presence of Lorentz violation would help to set a new upper bound on the magnitude of the Darwin-correction term, but establishing this lies beyond our present scope.


\section{Conclusions\label{sec:Conclusions}}

In this paper, we presented the modifications produced by the spontaneous
breaking of Lorentz symmetry over the Newtonian gravitational potential by means of the direct calculation of the scattering amplitude between two massive scalar particles interacting gravitationally.

First, we have introduced an action to the simplest gravity model
involving tensor fields, responsible for the spontaneous local Lorentz
breaking, coupled with the gravitational field. To construct the
gravitational sector for the minimal SME, we take a particular case of the so-called bumblebee model. After that, we separate the dynamical fields into the vacuum expectation values and the quantum fluctuations to analyze the effects of gravity-bumblebee coupling on the graviton dynamics. Inserting the solution of the equation of motion for the bumblebee field in the LV action, we have determined the modified 
kinetic term for the graviton field. Dealing these modifications in the form of a perturbative insertion, we have obtained a corrected propagator for which ghosts and tachyons are not present. 

As a result, we observed at first order in LV coupling $\xi$, an unconventional  spatial  dependence with respect to the  separation vector between the two bodies and which agrees with previous results obtained through post-Newtonian approximations for the gravitational sector of the SME. In second order in $\xi$, we verify the appearance of a Darwin-like correction term, independent of the VEV $b_{\mu}$ of the bumblebee field, reflecting the effect of the bumblebee fluctuation $\tilde{B}^{\mu}$ on the graviton propagation. This result corroborates the fact that at small distances where higher terms in the curvature are relevant, the gravitational force becomes much stronger and the local Lorentz symmetry might be violated. Moreover, a similar correction was obtained in a theory for the H\v orava-Lifshitz gravity  containing higher spatial derivatives \cite{ref25}.

Finally, a detailed analysis about the graviton spectrum corrections induced by the spontaneous Lorentz violation, in the context of bumblebee models, seems to be
a sensitive issue and is a subject for a forthcoming article.

\begin{acknowledgments}
The authors thank Professor Quentin G. Bailey for drawing our attention to some misconceptions in the preliminary version of this manuscript. We also are grateful to the referees for a number of helpful suggestions and comments. This work was partially supported by Coordena\c{c}\~ao de Aperfeiçoamento de
Pessoal de N\'ivel Superior (CAPES), Conselho Nacional de Pesquisas (CNPq), and
Funda\c{c}\~ao Cearense de apoio ao Desenvolvimento Cient\'ifico e Tecnol\'ogico
(FUNCAP).
\end{acknowledgments}


\begin{thebibliography}{99}
\bibitem{ref1} D. Mattingly, Living Rev. Rel. {\bf 8}, 5 (2005).

\bibitem{ref2} J. Alfaro, H. A. Morales-Tecotl and L. F. Urrutia, Phys. Rev. Lett. {\bf 84}, 2318 (2000); J. Alfaro, H. A. Morales-Tecotl and L. F. Urrutia, Phys. Rev. D {\bf 65}, 103509 (2002).

\bibitem{ref3} N. E. Mavromatos, Proc. Sci QG-PH, {\bf 027} (2007).

\bibitem{ref3.2} S. M. Carroll, J. A. Harvey, V. A. Kosteleck\'y, C. D. Lane, and T. Okamoto, Phys. Rev. Lett. {\bf 87}, 141601 (2001).

\bibitem{ref3.3} T. G. Rizzo, J. High Energy Phys. 09 (2005) 036; T. G. Rizzo, J. High Energy Phys. 11 (2010) 156.

\bibitem{ref3.4} V. Santos and C. A. S. Almeida, Phys. Lett. B \textbf{718},
1114 (2013).

\bibitem{ref3.5} M. Pospelov and Y. Shang, Phys. Rev. D {\bf 85}, 105001 (2012).

\bibitem{ref4} D. Colladay and V. A. Kosteleck\'y, Phys. Rev. D \textbf{55},
6760 (1997); Phys. Rev. D \textbf{58}, 116002 (1998).

\bibitem{refS} V. A. Kosteleck\'y and S. Samuel, Phys. Rev. D \textbf{39}, 683
(1989); {\bf Phys. Rev. D {\bf 40}, 1886 (1989)}.


\bibitem{ref4L} V. A. Kosteleck\'y and R. Lehnert, Phys. Rev. D {\bf 63},
065008 (2001).

\bibitem{ref5} V. A. Kosteleck\'y and M. Mewes, Phys. Rev. Lett. \textbf{87},
251304 (2001); Phys. Rev. D \textbf{66}, 056005 (2002); Phys. Rev. Lett.
\textbf{97}, 140401 (2006).

\bibitem{ref6} Q. G. Bailey and V. A. Kosteleck\'y, Phys. Rev. D {\bf 70},
076006 (2004); R. Casana, M.M. Ferreira Jr, C. E. H. Santos, Phys. Rev. D
{\bf 78}, 105014 (2008); R. Casana, M.M. Ferreira Jr, A. R. Gomes, P. R. D.
Pinheiro, Eur. Phys. J. C {\bf 62}, 573 (2009).

\bibitem{ref6B} R. Casana, M.M. Ferreira Jr, A. R. Gomes, P. R. D.
Pinheiro, Phys. Rev. D {\bf 80}, 125040 (2009); R. Casana, M.M. Ferreira Jr, A. R.
Gomes, F. E. P. dos Santos, Phys. Rev. D {\bf 82}, 125006 (2010); F.R. Klinkhamer,
M. Schreck, Nucl. Phys. {\bf B} {\bf 848}, 90 (2011); M. Schreck, Phys. Rev. D {\bf 86},
065038 (2012).

\bibitem{ref6C} F. R. Klinkhamer and M. Risse, Phys. Rev. D {\bf 77}, 016002
(2008); {\bf 77}, 117901 (2008); F. R. Klinkhamer and M. Schreck, Phys. Rev. D {\bf 78},
085026 (2008); B. Charneski, M. Gomes, T. Mariz, J. R. Nascimento and A. J.
da Silva, Phys. Rev. D \textbf{79}, 065007 (2009).

\bibitem{ref6D} B. Altschul, Nucl. Phys. {\bf B} {\bf 796}, 262 (2008); B.
Altschul, Phys. Rev. Lett. {\bf 98}, 041603 (2007); C. Kaufhold and F.R.
Klinkhamer, Phys. Rev. D {\bf 76}, 025024 (2007).

\bibitem{ref6E} V. A. Kostelecky and M. Mewes, Phys. Rev. D {\bf 80},
015020 (2009); M. Cambiaso, R. Lehnert, and R. Potting, Phys. Rev. D {\bf 85},
085023 (2012); M. Mewes, Phys. Rev. D {\bf 85}, 116012 (2012). B. Agostini, F. A.
Barone, F. E. Barone, P. Gaete, J. A. Helay\"el-Neto, Phys. Lett. B {\bf 708}, 212
(2012).

\bibitem{ref6F} R. C. Myers and M. Pospelov, Phys. Rev. Lett. {\bf 90},
211601 (2003); P.A. Bolokhov and M. Pospelov, Phys. Rev. D {\bf 77}, 025022
(2008); C.M. Reyes, L. F. Urrutia, and J. D. Vergara, Phys. Rev. D {\bf 78},
125011 (2008); C.M. Reyes, Phys. Rev. D {\bf 80}, 105008 (2009); {\bf 82}, 125036
(2010); J. Lopez-Sarrion and C. M. Reyes, Eur. Phys. J. C {\bf 72}, 2150 (2012);
C.M. Reyes, L. F. Urrutia, and J. D. Vergara, Phys. Lett. B {\bf 675}, 336 (2009).

\bibitem{ref7A} H. Belich, T. Costa-Soares, M. M. Ferreira Jr., J. A.
Helay\"el-Neto, Eur. Phys. J. C {\bf 41}, 421 (2005); H. Belich, T. Costa-Soares, M. M. Ferreira, Jr., J. A. Helay\"el-Neto, and M. T. D. Orlando, Phys. Lett. B
{\bf 639}, 675 (2006).

\bibitem{ref7B} H. Belich, E. O. Silva, M. M. Ferreira Jr. and M. T.
D. Orlando, Phys. Rev. D {\bf 83}, 125025 (2011); B. Charneski, M. Gomes, R. V.
Maluf and A. da Silva, Phys. Rev. D {\bf 86}, 045003 (2012); G. Gazzola, H. G.
Fargnoli, A. P. Baeta Scarpelli, M. Sampaio, and M. C. Nemes, J. Phys. G {\bf 39},
035002 (2012); A. P. Baeta Scarpelli, J. Phys. G {\bf 39}, 125001 (2012); E. O.
Silva and F. M. Andrade, EPL {\bf 101}, 51005 (2013).

\bibitem{ref8} R. Casana, M. M. Ferreira, R. V. Maluf and F. E. P.
dos Santos, Phys. Rev. D {\bf 86}, 125033 (2012); R. Casana, M. M. Ferreira, Jr.,
E. Passos, F. E. P. dos Santos and E. O. Silva, Phys. Rev. D {\bf 87}, 047701
(2013); R. Casana, M. M. Ferreira Jr, R. V. Maluf, F. E. P. dos Santos,
arXiv:1302.2375 [hep-th].


\bibitem{ref14} J. W. Moffat, Int. J. Mod. Phys. D \textbf{12}, 1279 (2003);
V. A. Kosteleck\'y, Phys. Rev. D \textbf{69}, 105009 (2004).

\bibitem{ref15} R. Bluhm and V. A. Kosteleck\'y, Phys. Rev. D \textbf{71},
065008 (2005).

\bibitem{ref16} Q. G. Bailey and V. A. Kosteleck\'y, Phys. Rev. D \textbf{74}
, 045001 (2006).

\bibitem{ref17} R. Bluhm, Shu-Hong Fung and V. A. Kosteleck\'y, Phys. Rev. D
\textbf{77}, 065020 (2008).

\bibitem{ref18} V. A. Kosteleck\'y and R. Potting, Phys. Rev. D \textbf{79},
065018 (2009).

\bibitem{ref19} B. Altschul, Q. G. Bailey and V. A. Kosteleck\'y, Phys. Rev.
D \textbf{81}, 065028 (2010).

\bibitem{ref19.2} A. F. Ferrari, M. Gomes, J. R. Nascimento, E. Passos, A. Yu. Petrov, A. J. da Silva, Phys. Lett. B {\bf 652}, 174 (2007).

\bibitem{ref20} V. A. Kosteleck\'y and J. D. Tasson, Phys. Rev. D \textbf{83}
, 016013 (2011).

\bibitem{ref20.1} R. Tso and Q. G. Bailey, Phys. Rev. D {\bf 84}, 085025 (2011).  

\bibitem{Sakurai} J. J. Sakurai,  {\it Advanced Quantum Mechanics} (Addison-Wesley, Reading, MA, 1967), p. 86.

\bibitem{ref22} W. F. Chen and G. Kunstatter, Phys. Rev. D {\bf 62},
105029 (2000); C. D. Carone, M. Sher and M. Vanderhaeghen, Phys. Rev. D {\bf 74},
077901 (2006); A. Ferrero and B. Altschul, Phys. Rev. D \textbf{84}, 065030
(2011).


\bibitem{ref23} J. F. Donoghue, Phys. Rev. D \textbf{50}, 3874 (1994).

\bibitem{refH} J. L. Boldo, J. A. Helay\"el-Neto, L.M. de Moraes,
C. A. G. Sasaki, V. J. V. Otoya, Phys. Lett. B {\bf 689}, 112 (2010); C. A Hernaski,
A. A. Vargas-Paredes, J. A. Helay\"el-Neto, Phys. Rev. D {\bf 80}, 124012 (2009).

\bibitem{ref24} N. E. J. Bjerrum-Bohr, J. F. Donoghue and B. R. Holstein,
Phys. Rev. D \textbf{67}, 084033 (2003).

\bibitem{ref28} Floyd W. Stecker, Astroparticle Phys. \textbf{35}, 95
(2011); V. A. Kosteleck\'y, N. Russell, Rev. Mod. Phys. \textbf{83}, 11
(2011);  V. A. Kosteleck\'y, R. Lehnert, and M. Perry, Phys. Rev. D
\textbf{68}, 123511 (2003); R. Bluhm, Lect. Notes Phys. \textbf{702}, 191 (2006).


\bibitem{ref29} D. E. Lebach, B.E. Corey, I. I. Shapiro, M. I. Ratner, J. C.
Webber, A. E. E. Rogers, J. L. Davis, and T. A. Herring, Phys. Rev. Lett.
\textbf{75}, 1439 (1995); B. Bertotti, L. Iess, and P. Tortora, Nature (London) {\bf 425}, 374 (2003); S. S. Shapiro, J. L. Davis, D. E. Lebach, and J.
S. Gregory, Phys. Rev. Lett. \textbf{92},121101 (2004).

\bibitem{Donoghue}  J. F. Donoghue, Phys. Rev. Lett. {\bf 72}, 2996 (1994); Proc. VIII M. Grossmann Conf. on General Relativity, 	arXiv:gr-qc/9712070.

\bibitem{Accioly} A. Accioly, H. Blas, Mod. Phys. Lett. A {\bf22}, 961 (2007).



\bibitem{Yang} S.-Q. Yang, B.-F. Zhan, Q.-L. Wang, C.-G. Shao, L.-C. Tu, W.-H. Tan and J. Luo, Phys. Rev. Lett. {\bf 108}, 081101 (2012); C. D. Hoyle, D. J. Kapner, B. R. Heckel, E. G. Adelberger, J. H. Gundlach, U. Schmidt and H. E. Swanson, Phys. Rev. D {\bf 70}, 042004 (2004). 



\bibitem{ref25} F. S. Bemfica and M. Gomes, Phys. Rev. D \textbf{84}, 084022
(2011).

\end{thebibliography}
\end{document}